\documentclass[aps, prd, superscriptaddress, preprintnumbers,showkeys]{revtex4}

\usepackage{graphicx,amsfonts,amsmath,amssymb,amstext}
\usepackage{float,wrapfig}
\usepackage{subfigure, psfrag}
\usepackage{dsfont}
\usepackage{color}
\usepackage{bm}

\newcommand{\be}{\begin{equation}}
\newcommand{\ee}{\end{equation}}
\newcommand{\ba}{\begin{eqnarray}}
\newcommand{\ea}{\end{eqnarray}}

\definecolor{purple}{rgb}{0.8,0,0.6}
\definecolor{darkgreen}{rgb}{0.00,0.6,0.00}

\begin{document}

\title{Electronic properties of strained double-Weyl systems}
\date{August 16, 2018}

\author{P.~O.~Sukhachov}
\email{psukhach@uwo.ca}
\affiliation{Department of Applied Mathematics, Western University, London, Ontario, Canada N6A 5B7}

\author{E.~V.~Gorbar}
\affiliation{Department of Physics, Taras Shevchenko National Kiev University, Kiev, 03680, Ukraine}
\affiliation{Bogolyubov Institute for Theoretical Physics, Kiev, 03680, Ukraine}

\author{I.~A.~Shovkovy}
\affiliation{College of Integrative Sciences and Arts, Arizona State University, Mesa, Arizona 85212, USA}
\affiliation{Department of Physics, Arizona State University, Tempe, Arizona 85287, USA}

\author{V.~A.~Miransky}
\affiliation{Department of Applied Mathematics, Western University, London, Ontario, Canada N6A 5B7}

\begin{abstract}
The effects of strains on the low-energy electronic properties of double-Weyl phases are studied in solids and cold-atom optical lattices. The principal finding is that deformations do not couple, in general, to the low-energy effective Hamiltonian as a pseudoelectromagnetic gauge potential. The response of an optical lattice to strains is simpler, but still only one of several strain-induced terms in the corresponding low-energy Hamiltonian can be interpreted as a gauge potential. Most interestingly, the strains can induce a nematic order parameter that splits a double-Weyl node into a pair of Weyl nodes with the unit topological charges. The effects of deformations on the motion of wavepackets in the double-Weyl optical lattice model are studied. It is found that, even in the undeformed lattices, the wavepackets with opposite topological charges can be spatially split. Strains, however, modify their velocities in a very different way and lead to a spin polarization of the wavepackets.
\end{abstract}


\maketitle

\section{Introduction}
\label{sec:Introduction}

In recent years, Weyl semimetals attracted a significant attention of the condensed matter community.
Such an interest is connected with the fact that the low-energy dynamics of their quasiparticles in the
vicinity of band-touching points (Weyl nodes) is described by the three-dimensional (3D)
relativisticlike Weyl equation (for recent reviews, see Refs.~\cite{Yan-Felser:2017-Rev,Hasan-Huang:2017-Rev,Armitage-Vishwanath:2017-Rev}).
Weyl nodes are separated in momentum by $2\mathbf{b}$ and/or in energy by $2b_0$ leading
to the breakdown of the time-reversal (TR) and/or parity-inversion symmetries, respectively.
Because of their relativisticlike low-energy quasiparticles, Weyl semimetals may reveal
a variety of seemingly high-energy phenomena, such as those triggered by the celebrated chiral
anomaly~\cite{Adler,Bell-Jackiw}. Weyl materials also possess nontrivial topological
properties that include the monopolelike Berry curvature \cite{Berry:1984}, unconventional open
surface states known as the Fermi arcs \cite{Savrasov,Aji,Haldane}, etc.
As was shown by Nielsen and Ninomiya \cite{Nielsen-Ninomiya-1,Nielsen-Ninomiya-2}, because of their nontrivial topology,
the Weyl nodes in crystals always come in pairs of opposite chirality or, equivalently, topological charge.
In addition, these material have unconventional transport properties, in particular, the ``negative" longitudinal magnetoresistivity
(i.e., the resistivity in the direction of the magnetic field decreases with the field's strength)
predicted in Ref.~\cite{Nielsen}. (For recent reviews of transport phenomena in Weyl semimetals,
see Refs.~\cite{Lu-Shen-rev:2017,Wang-Lin-rev:2017,Gorbar:2017lnp}.)

In some materials, Weyl nodes of equal topological charges can merge at the same points in the Brillouin zone and
produce the multi-Weyl nodes whose topological charges $n_{\text{\tiny W}}$ are greater than one.
The dispersion relations in the vicinity of such nodes are described by higher than linear dependencies on momenta in two directions.
It is shown in Ref.~\cite{Fang-Bernevig:2012} that crystallographic
point symmetries can protect only the multi-Weyl nodes with $|n_{\text{\tiny W}}|\leq3$.
The realization of the corresponding multi-Weyl semimetals was theoretically proposed in $\mbox{HgCr}_2\mbox{Se}_4$~\cite{Xu-Fang:2011,Fang-Bernevig:2012}
and $\mbox{SrSi}_2$~\cite{Huang}. Multi-Weyl semimetals inherit almost all nontrivial features
of usual Weyl semimetals.
In particular, numerical calculations suggest the presence of multiple surface Fermi arcs~\cite{Xu-Fang:2011,Fang-Bernevig:2012,Huang}. Also, many of their anomalous transport coefficients are predicted to
be the same as in usual Weyl semimetals \cite{Huang:2017rpa} up to the factor $n_{\text{\tiny W}}$.

Remarkably, the realization of the Weyl phases is not limited only to solids. It was shown
that the Weyl equation could describe the low-energy dynamics of ultracold atoms in 3D
optical lattices \cite{Jiang:2012,Ganeshan-DasSarma:2015,Dubcek-Buljan:2015,Li-Sarma:2015},
electromagnetic waves in photonic crystals \cite{Chen-Chan:2016}, and even sound waves in
special heterostructures \cite{Xiao-Chan:2015,Liu-Xia:2018}. Due to their great
tunability, optical lattices allow also for the realization of the double-Weyl phase
\cite{Lepori-Burrello:2016,Mai-Zhu:2017}. In addition, it was proposed that one can simulate the double-Weyl semimetals with synthetic non-Abelian SU(2) gauge potentials in such systems \cite{Lepori-Burrello:2016}.
While, to the best of our knowledge, there are currently no experimental observations of the double-Weyl phase in solids, the cold-atom systems could provide an alternative platform to study the properties of this topologically nontrivial phase.
Indeed, the experimental observation of the double-Weyl nodes was already reported in photonic crystals \cite{Chen-Chan:2016} and acoustic semimetals \cite{Liu-Xia:2018}.

Although Weyl phases show many qualitative properties of truly relativistic matter,
they also allow for phenomena that are rather uncommon  in high-energy physics.
In particular, the generation of the axial gauge potential $\mathbf{A}_5$ is
rather exotic in high-energy systems.
However, it was shown
that $\mathbf{A}_5$ can be relatively easily generated by mechanical strains in Weyl semimetals~\cite{Zhou:2012ix,Zubkov:2015,Cortijo:2015jja,Cortijo:2016yph,Cortijo:2016,Pikulin:2016,Grushin-Vishwanath:2016,Liu-Pikulin:2016,Arjona:2018ryu} or by deformations in the corresponding optical lattices \cite{Roy-Grushin:2018}.
This gauge potential effectively captures the corrections to the kinetic energy of quasiparticles caused by strain-induced modifications of hopping parameters.
Further, unlike the usual electromagnetic gauge potential $\mathbf{A}$, its axial counterpart is observable.
Indeed, it can be interpreted as a time and coordinate dependent separation between the Weyl nodes.
Unlike ordinary electric $\mathbf{E}$ and magnetic $\mathbf{B}$ fields, their pseudoelectromagnetic counterparts interact with the fermions of opposite chirality (or topological charge) with different signs.
It is predicted that the strain-induced pseudoelectromagnetic fields lead to various effects, including the strain-enhanced conductivity
\cite{Pikulin:2016,Grushin-Vishwanath:2016}, the ultrasonic attenuation \cite{Pikulin:2016}, the electromagnetic emission \cite{Pikulin:2016},
and the quantum oscillations without magnetic fields \cite{Liu-Pikulin:2016}.
Further, the Fermi arcs can be reinterpreted as the zeroth pseudo-Landau levels due to the pseudomagnetic field $\mathbf{B}_5$ localized at the boundary \cite{Grushin-Vishwanath:2016}.

To the best of our knowledge, however, all existing studies of strain effects considered only
usual Weyl semimetals with the topological charges of the Weyl nodes $n_{\text{\tiny W}}=\pm 1$.
Since the corresponding low-energy Hamiltonian $\bm{\sigma}\cdot\mathbf{k}$ is linear in the wave vector $\mathbf{k}$, all $\mathbf{k}$-independent perturbations, except those proportional to the unit matrix, can only shift the positions of the Weyl nodes and, therefore, are naturally interpreted in terms of an axial gauge potential.
Its axial character is evident from the fact that static strains do not break the TR symmetry.
It is worth noting that, in addition to the axial gauge potential, there might be other terms related to the tilt of Weyl cones, the anisotropic Fermi velocity, and the pseudo-Zeeman term \cite{Arjona:2018ryu}. However, these additional terms depend on wave vector and are not the main focus of this study.
Further, the low-energy Hamiltonians of multi-Weyl phases are not linear in $\mathbf{k}$ \cite{Fang-Bernevig:2012}.
Therefore, it is not obvious whether strains in double-Weyl systems ($n_{\text{\tiny W}}=\pm2$) couple in the same way as in usual ones ($n_{\text{\tiny W}}=\pm1$). Indeed, while in Refs.~\cite{Fang-Bernevig:2012,Huang,Mai-Zhu:2017} the symmetry-based arguments were used to show that a double-Weyl node can be split by strains into two usual Weyl nodes, no detailed analysis was provided.
The main goal of this paper is to show how strains affect the low-energy electronic properties of double-Weyl phases and to
determine possible observable effects. The latter include the nontrivial motion of the wavepackets, whose propagation becomes inhibited depending on chirality.
Note that the motion of wavepackets was also studied in the optical lattices with usual Weyl nodes in Ref.~\cite{Roy-Grushin:2018}, where pseudoelectromagnetic fields were introduced phenomenologically.

The paper is organized as follows. A solid-state model is introduced and the effects of strains are studied in Sec.~\ref{sec:model-solid}.
The results for ultracold atoms in an optical lattice are given in Sec.~\ref{sec:Model-optical}.
Sec.~\ref{sec:wavepackets} is devoted to the motion of wavepackets in the deformed optical lattice model.
The results are discussed and summarized in Sec.~\ref{sec:Summary}.
Some technical details, including the results for the nonzero components of the Gr\"{u}neisen tensors and the Fourier transforms of the strained lattice Hamiltonians, are presented in Appendix \ref{sec:Gruneisen-H-app}.

\section{Solid state model}
\label{sec:model-solid}

\subsection{General formulation and lattice Hamiltonian}
\label{sec:model-solid-general}

Let us start our study of the strain effects in a double-Weyl semimetal.
We employ the following effective model that describes the low-energy dynamics of HgCr$_2$Se$_4$ \cite{Xu-Fang:2011}:
\begin{equation}
\label{model-H-eff}
H_{\rm eff} = \left(
                \begin{array}{cc}
                  M_0 -\beta k^2 & D k_z k_{-}^2 \\
                  D k_z k_{+}^2 & -\left(M_0 -\beta k^2\right) \\
                \end{array}
              \right).
\end{equation}
This Hamiltonian was obtained from the first-principles calculations and is written in the basis of the $P$ and $S$ states, such as $\left|\frac{3}{2},\frac{3}{2} \right\rangle$ and $\left|S,-\frac{1}{2} \right\rangle$.
(Here the first and second numbers denote the total angular momentum and its projection, respectively.)
Note, however, that these states are, in fact, nontrivial combinations of the $\left|s\right\rangle$, $\left|p\right\rangle$, and
$\left|d\right\rangle$ orbitals of Se, Hg, and Cr atoms.
Further, $M_0>0$, $\beta>0$, and $D$ are model parameters. In addition, $k_{\pm}=k_x\pm i k_y$ and $k=|\mathbf{k}|$.

\begin{figure}[t]
\begin{center}
\includegraphics[width=1\textwidth]{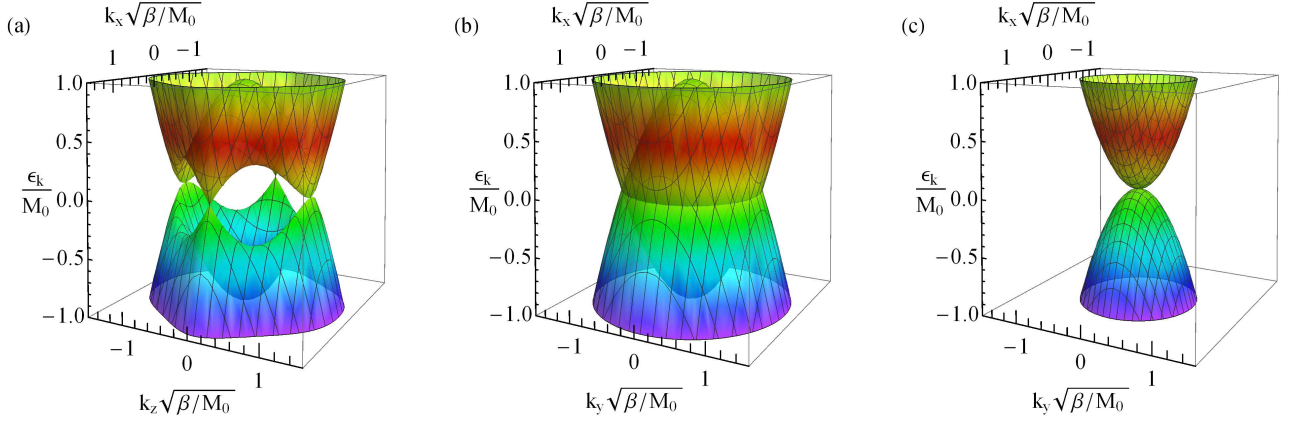}
\caption{The energy dispersion relation (\ref{model-energy}) at $k_y=0$ (panel a), $k_z=0$ (panel b), and $k_z=\sqrt{M_0/\beta}$ (panel c) plotted for
$D^2M_0/\beta^3=1$.}
\label{fig:model-energy-3D}
\end{center}
\end{figure}

The energy spectrum of Hamiltonian (\ref{model-H-eff}) reads
\begin{equation}
\label{model-energy}
\epsilon_{\mathbf{k}} = \pm \sqrt{\left(M_0 -\beta k^2\right)^2+D^2k_z^2k_{\perp}^4},
\end{equation}
where $k_{\perp}=\sqrt{k_x^2+k_y^2}$. We plot the above energy dispersion for several values of momenta in
Fig.~\ref{fig:model-energy-3D}.
As one can see from  Fig.~\ref{fig:model-energy-3D}(a), such a spectrum contains two types of gapless features.
The first type is a ring-like intersection located at $k_{\perp}^2=M_0/\beta$ and $k_z=0$.
It is shown in Fig.~\ref{fig:model-energy-3D}(b) at $k_z=0$ and corresponds to the two touching points in Fig.~\ref{fig:model-energy-3D}(a) at $k_z=0$. What is important for us, there are also two double-Weyl nodes at $k_z=\pm \sqrt{M_0/\beta}$.
We show one of them at $k_{z}=\sqrt{M_0/\beta}$  in Fig.~\ref{fig:model-energy-3D}(c).

In the vicinity of the double-Weyl nodes with the topological charges $n_{\text{\tiny W}}=\pm2$, i.e., at $\mathbf{k}=\left\{\delta k_x, \delta k_y, \delta k_z +n_{\text{\tiny W}}\sqrt{M_0/4\beta}\right\}$, Hamiltonian (\ref{model-H-eff}) reads
\begin{equation}
\label{model-H-eff-chiral}
H_{\rm eff} \approx  -\sigma_z\left(n_{\text{\tiny W}} \sqrt{M_0 \beta} \delta k_z +\beta \delta k^2_{\perp} \right) + \frac{n_{\text{\tiny W}} D}{2} \sqrt{\frac{M_0}{\beta}}\left( \sigma_{+} \delta k_{-}^2 +\sigma_{-} \delta k_{+}^2\right) +O(\delta k_z^2, \delta k_z \delta k_{\perp}^2),
\end{equation}
where $\delta k_{\pm}=\delta k_{x}\pm i\delta k_{y}$, $\delta k^2_{\perp}=\delta k_x^2+\delta k_y^2$, and $\sigma_{\pm}=(\sigma_x \pm i\sigma_y)/2$ are combinations of the Pauli matrices that act in the space of the $P$ and $S$ states.
The low-energy Hamiltonian in Eq.~(\ref{model-H-eff-chiral}) represents a minimal model of the double-Weyl semimetal with the additional ring-like feature.

In order to study the effects of elastic deformations, one should use a lattice model in the coordinate space rather than the effective low-energy model defined by Eq.~(\ref{model-H-eff}).
Therefore, by employing the approach in Ref.~\cite{Shen:2017-book} and assuming a hypercubic structure of the lattice, we obtain the following tight-binding analog of the effective model (\ref{model-H-eff}):
\begin{eqnarray}
\label{model-strains-2-H-r}
H_{\rm latt}&=& \sigma_z \left(M_0- \frac{6\beta}{a^2}\right) \sum_{\mathbf{r}} c^{\dag}_{\mathbf{r}}c_{\mathbf{r}} + \sigma_z \frac{\beta}{a^2} \sum_{\mathbf{r}} \sum_{j=x,y,z} \left(c^{\dag}_{\mathbf{r}} c_{\mathbf{r}+\mathbf{a}_j} +c^{\dag}_{\mathbf{r}+\mathbf{a}_j} c_{\mathbf{r}}\right) \nonumber\\
&-& \sigma_x  \frac{iD}{2a^3} \sum_{\mathbf{r}} \sum_{j=1}^{4} (-1)^{j+1} \left(c^{\dag}_{\mathbf{r}} c_{\mathbf{r}+\mathbf{a}_j^{\prime}} -c^{\dag}_{\mathbf{r}+\mathbf{a}_j^{\prime}} c_{\mathbf{r}}\right) +\sigma_y \frac{iD}{4a^3} \sum_{\mathbf{r}} \sum_{j=1}^4 \left(2\delta_{j1}-1\right) \left(c^{\dag}_{\mathbf{r}} c_{\mathbf{r}+\mathbf{a}_j^{\prime \prime}} -c^{\dag}_{\mathbf{r}+\mathbf{a}_j^{\prime \prime}} c_{\mathbf{r}}\right) .
\end{eqnarray}
Here $c^{\dag}_{\mathbf{r}}$ and $c_{\mathbf{r}}$ denote the creation and annihilation operators at
position $\mathbf{r}$ and $a$ is the lattice constant.
The hopping terms for the transitions between the same types of states (e.g., $S \leftrightarrow S$) are represented
by nearest-neighbour hoppings that are described by the three basis vectors of the
hypercubic lattice, $\mathbf{a}_j=a\hat{\mathbf{j}}$, where $\hat{\mathbf{j}}$ denotes the unit vector
in the direction $j=x,y,z$. It is important to note that the transitions between the different
types of states (e.g., $S \leftrightarrow P$) in Hamiltonian (\ref{model-strains-2-H-r}) are not
restricted only to the nearest neighbors (indeed, $|\mathbf{a}^{\prime}_j|$ and
$|\mathbf{a}^{\prime\prime}_j|$ exceed $|\mathbf{a}_j|=a$ by $\sqrt{2}$ and $\sqrt{3}$ times, respectively).
The corresponding vectors are defined as follows:
\begin{eqnarray}
\label{model-lattice-real-a1-1}
\mathbf{a}_1^{\prime}&=&a\left\{0,1,1\right\},\quad\quad \mathbf{a}_1^{\prime \prime}=a\left\{1,1,1\right\},\\
\label{model-lattice-real-a1-2}
\mathbf{a}_2^{\prime}&=&a\left\{1,0,1\right\},\quad\quad \mathbf{a}_2^{\prime \prime}=a\left\{-1,1,1\right\},\\
\label{model-lattice-real-a1-3}
\mathbf{a}_3^{\prime}&=&a\left\{0,-1,1\right\},\quad\quad \mathbf{a}_3^{\prime \prime}=a\left\{1,-1,1\right\},\\
\label{model-lattice-real-a1-4}
\mathbf{a}_4^{\prime}&=&a\left\{-1,0,1\right\},\quad\quad \mathbf{a}_4^{\prime \prime}=a\left\{1,1,-1\right\}.
\end{eqnarray}

By making use of the tight-binding model (\ref{model-strains-2-H-r}) and performing the Fourier transform, we obtain
\begin{equation}
\label{model-strains-2-H-k-0}
H_{\rm latt}= \sum_{\mathbf{k}} c^{\dag}_{\mathbf{k}}\mathcal{H}(\mathbf{k})c_{\mathbf{k}},
\end{equation}
where
\begin{equation}
\label{model-strains-2-H-k}
\mathcal{H}(\mathbf{k}) = \sigma_z \left(M_0-  \frac{6\beta}{a^2}\right)
+ \sigma_z \frac{2\beta}{a^2} \sum_{j=x,y,z} \cos{(\mathbf{k} \cdot\mathbf{a}_j)}
+\sigma_x \frac{D}{a^3} \sum_{j=1}^4 (-1)^{j+1} \sin{(\mathbf{k} \cdot\mathbf{a}_j^{\prime})}
- \sigma_y \frac{D}{2a^3} \sum_{j=1}^4 \left(2 \delta_{j1}-1\right) \sin{(\mathbf{k} \cdot\mathbf{a}_j^{\prime \prime})}.
\end{equation}
One can easily check that the effective Hamiltonian (\ref{model-H-eff}) is reproduced in the continuous limit $a\to0$.

\subsection{Strains and their effects on the low-energy effective Hamiltonian}
\label{sec:model-strains-2}

As in Weyl semimetals with the unit topological charge~\cite{Shapourian-Ryu:2015,Cortijo:2015jja,Cortijo:2016yph}, strains are
included via the change of hopping parameters. To the linear order in deformations, a general expression for the tight-binding
parameter
$t\left(\mathbf{a}_j+\delta \mathbf{r}(\mathbf{a}_j)\right)$ that describes a hopping along the $\mathbf{a}_j$ direction
is given by
\begin{eqnarray}
\label{model-strains-2-t-general}
t \left(\mathbf{a}_j+\delta \mathbf{r}(\mathbf{a}_j)\right) &\approx& t(\mathbf{a}_j)\left[1 - \sum_{i=x,y,z} \beta_{ij}^{\rm G}
\frac{\left(\delta \mathbf{r}(\mathbf{a}_j) \cdot \mathbf{a}_i\right)}{|\mathbf{a}_i|}\right] +O(\delta r^2).
\end{eqnarray}
Here $\beta^{\rm G}_{ij}$ is the tensor that relates the changes of the hopping amplitude to deformations.
Therefore, in what follows, we will call such a structure the Gr\"{u}neisen tensor.
(Strictly speaking, the analogy with the Gr\"{u}neisen parameter is not exact because $\beta^{\rm G}_{ij}$ is a tensor and it is not dimensionless.)
The modification of the hopping length $\delta \mathbf{r}(\mathbf{a}_j)$ can be
expressed through the displacement vector $\mathbf{u}$ as follows:
\begin{equation}
\label{model-strains-2-dr}
\delta \mathbf{r}(\mathbf{a}_j) = \left(\mathbf{a}_j\cdot\bm{\nabla}\right) \mathbf{u}.
\end{equation}
Henceforth, it is convenient to use the unsymmetrized strain tensor $\hat{u}$ whose components are defined as {$u_{ij}=\partial_{i}u_j$, where $i,j=x,y,z$.

By taking into account the matrix structure of model (\ref{model-strains-2-H-r}) the strains connected with the $S\leftrightarrow S$ and $P \leftrightarrow P$ hoppings can be described via the following replacements:
\begin{eqnarray}
\label{model-strains-2-tssy}
\frac{\beta}{a^2} \frac{1+\sigma_z}{2} &\to& \frac{\beta}{a^2} \frac{1+\sigma_z}{2}\left[1- \sum_{i=x,y,z} \beta_{ij}^{(SS)} \frac{\left(\delta \mathbf{r}(\mathbf{a}_j) \cdot \mathbf{a}_i\right)}{|\mathbf{a}_i|} \right],\\
\label{model-strains-2-tppy}
\frac{\beta}{a^2} \frac{1-\sigma_z}{2} &\to& \frac{\beta}{a^2} \frac{1-\sigma_z}{2}\left[1 -  \sum_{i=x,y,z} \beta_{ij}^{(PP)} \frac{\left(\delta \mathbf{r}(\mathbf{a}_j) \cdot \mathbf{a}_i\right)}{|\mathbf{a}_i|}\right].
\end{eqnarray}
Further, the hoppings between the $S$ and $P$ states are
\begin{eqnarray}
\label{model-strains-2-tspx}
-\frac{D}{2a^3}(-1)^{j+1}\sigma_x &\to& -\frac{D}{2a^3}\sigma_x\left[(-1)^{j+1}-\sum_{i=x,y,z} \beta_{ij}^{(x)} \frac{\left(\delta \mathbf{r}(\mathbf{a}_j^{\prime}) \cdot \mathbf{a}_i\right)}{|\mathbf{a}_i|}\right],\\
\label{model-strains-2-tspy}
\frac{D}{4a^3} \left(2\delta_{j1}-1\right) \sigma_y &\to& \frac{D}{4a^3}\sigma_y \left[\left(2\delta_{j1}-1\right) - \sum_{i=x,y,z} \beta_{ij}^{(y)} \frac{\left(\delta \mathbf{r}(\mathbf{a}_j^{\prime \prime}) \cdot \mathbf{a}_i\right)}{|\mathbf{a}_i|}\right].
\end{eqnarray}

Since the effective Hamiltonian (\ref{model-H-eff}) possesses a $C_4$ symmetry with respect to the $z$ axis, we employ the same symmetry constraints for the Gr\"{u}neisen tensor.
The $C_4$ symmetry of the corresponding Fourier transform is defined in the standard way
\begin{equation}
\label{model-strains-2-symmetry-H-C4}
C \mathcal{H}(\mathbf{k}) C^{-1} = \mathcal{H}(R_4\mathbf{k}).
\end{equation}
Here $R_4$ is the standard rotation operator with respect to the $z$ axis and $C=\sigma_z$ \cite{Fang-Bernevig:2012}.
Taking into account the $C_4$ symmetry, it is possible to significantly reduce the number of the nonzero components of the Gr\"{u}neisen tensors.
The corresponding results are given by Eqs.~(\ref{model-strains-2-symmetry-beta-nonzero-be})--(\ref{model-strains-2-symmetry-beta-nonzero-ee}) in Appendix \ref{sec:Gruneisen-app}.
Here, we note that the independent components are $\beta_{xx}^{(A)}$, $\beta_{xy}^{(A)}$, and $\beta_{zz}^{(A)}$, where $A=SS,PP$, as well as $\beta_{x1}^{(B)}$, $\beta_{x2}^{(B)}$, and $\beta_{z1}^{(B)}$, where $B=x,y$.

Due to its bulky form, the Fourier transform of the strained lattice Hamiltonian is given by Eq.~(\ref{model-strains-2-H-k-strain}) in Appendix \ref{sec:H-app}. Here, similarly to Eq.~(\ref{model-H-eff-chiral}), it is sufficient to expand it near the double-Weyl nodes and assume the continuous limit. The resulting Hamiltonian reads
\begin{eqnarray}
\label{results-revised-H-strain-1}
\mathcal{H}_{\rm strain}(\mathbf{k}) &\approx&V_0 - \sigma_z\left[n_{\text{\tiny W}} \sqrt{M_0 \beta} \left(\delta k_z-eA_z^{(z)}\right) +\beta \delta k_{\perp}^2
\right]\nonumber\\
&+& \sigma_x \left\{\frac{n_{\text{\tiny W}} D}{2} \sqrt{\frac{M_0}{\beta}} \left[\left(\delta k_x-eA_x^{(x)}\right)^2-\left(\delta k_y-eA_y^{(x)}\right)^2\right] +V_x \right\}
\nonumber\\
&+& \sigma_y \left[n_{\text{\tiny W}} D \sqrt{\frac{M_0}{\beta}} \left(\delta k_x-eA_x^{(y)}\right) \left(\delta k_y-eA_y^{(y)}\right)
+V_y
\right] + O(\delta k_z^2, \delta k_z\delta k_{\perp}^2,\delta k_z\hat{u}, \delta k_{\perp}^2\hat{u}),
\end{eqnarray}
where
\begin{equation}
\label{results-revised-V0-1}
V_0 = -\frac{\beta}{a} \left[\left(\beta_{xx}^{(SS)}-\beta_{xx}^{(PP)}\right)\left(u_{xx}+u_{yy}\right) +\left(\beta_{xy}^{(SS)}-\beta_{xy}^{(PP)}\right) \left(u_{yx}-u_{xy}\right) +\left(\beta_{zz}^{(SS)}-\beta_{zz}^{(PP)}\right)u_{zz}\right]
\end{equation}
denotes the strain-induced scalar potential term and
\begin{equation}
\label{results-revised-Azz-1}
A_z^{(z)} = -\frac{1}{n_{\text{\tiny W}}ea} \sqrt{\frac{\beta}{M_0}} \left[\left(\beta_{xx}^{(SS)}+\beta_{xx}^{(PP)}\right)\left(u_{xx}+u_{yy}\right) +\left(\beta_{xy}^{(SS)}+\beta_{xy}^{(PP)}\right) \left(u_{yx}-u_{xy}\right) +\left(\beta_{zz}^{(SS)}+\beta_{zz}^{(PP)}\right)u_{zz}\right]
\end{equation}
is the $z$ component of a strain-induced gauge potential. Strains affect the non-diagonal terms rather nontrivially and the
corresponding corrections in the terms at the $\sigma_x$ and $\sigma_y$ matrices read
\begin{eqnarray}
\label{results-revised-Axx-1}
A_x^{(x)} &=& \frac{2}{n_{\text{\tiny W}}ea} \sqrt{\frac{\beta}{M_0}} \left[\beta_{x1}^{(x)} u_{zy} +\beta_{x2}^{(x)} u_{zx} -\beta_{z3}^{(x)} u_{xz}\right],\\
\label{results-revised-Ayx-1}
A_y^{(x)} &=& -\frac{2}{n_{\text{\tiny W}}ea} \sqrt{\frac{\beta}{M_0}} \left[\beta_{x1}^{(x)} u_{zx} -\beta_{x2}^{(x)} u_{zy} +\beta_{z3}^{(x)} u_{yz}\right],\\
\label{results-revised-Dzx-1}
V_x &=& \frac{n_{\text{\tiny W}} D}{a} \sqrt{\frac{M_0}{\beta}} \left[\beta_{x1}^{(x)} \left(u_{xy}+u_{yx}\right) +\beta_{x2}^{(x)} \left(u_{xx}-u_{yy}\right)\right]
\end{eqnarray}
and
\begin{eqnarray}
\label{results-revised-Axy-1}
A_x^{(y)} &=& \frac{1}{n_{\text{\tiny W}}ea} \sqrt{\frac{\beta}{M_0}} \left[ \beta_{x1}^{(y)} \left(u_{zy}-u_{zx}\right) - \beta_{x2}^{(y)} \left(u_{zy}+u_{zx}\right) +2\beta_{z3}^{(y)} u_{xz}\right],\\
\label{results-revised-Ayy-1}
A_y^{(y)} &=& -\frac{1}{n_{\text{\tiny W}}ea} \sqrt{\frac{\beta}{M_0}} \left[ \beta_{x1}^{(y)} \left(u_{zx}+u_{zy}\right) - \beta_{x2}^{(y)} \left(u_{zx}-u_{zy}\right) -2\beta_{z3}^{(y)} u_{yz}\right],\\
\label{results-revised-Dzy-1}
V_y &=& -\frac{n_{\text{\tiny W}} D}{2a} \sqrt{\frac{M_0}{\beta}} \left[\beta_{x1}^{(y)} \left(u_{xx} +u_{xy}-u_{yx} -u_{yy}\right) -\beta_{x2}^{(y)} \left(u_{xx} -u_{xy}-u_{yx} -u_{yy}\right)\right],
\end{eqnarray}
respectively. Note that we assumed that the strain-induced terms $\propto \partial_iu_j$ can be treated as weak spatial variations of parameters in the momentum-space Hamiltonian. Further, both deformations and deviations of momenta from the double-Weyl nodes are small. Therefore, all higher-order terms, i.e., $O(\delta k_z\hat{u}, \delta k_{\perp}^2\hat{u})$, were neglected.

As one can clearly see from the strained Hamiltonian (\ref{results-revised-H-strain-1}), there are significant
modifications due to deformations. More importantly, they cannot be generally described by
a single gauge potential. Indeed, while some terms can be interpreted as the components of
strain-induced gauge potentials, their form is nonuniversal. For example, the diagonal terms with $\delta k^2_{\perp}$
in Hamiltonian (\ref{results-revised-H-strain-1}) contain neither $x$ nor $y$ components of the gauge
potential.
Further, while the strain-induced corrections in the off-diagonal parts of the Hamiltonian look like axial gauge potentials, they are different and, what is crucial, cannot be described in such a way.
We also found that there is the scalar potential term $V_0$ induced by the
difference between the Gr\"{u}neisen tensors for the $S\leftrightarrow S$ and $P\leftrightarrow P$ transitions.
In addition, deformations lead to new terms $V_x$ and $V_y$, which, as we will show in Subsec.~\ref{sec:strains-optical} for a double-Weyl lattice model,
could play the role of
nematic order parameters. Thus, we conclude that strains in the solid-state model of double-Weyl systems can be
described in terms of a gauge potential at best only in some special cases.

At the end of this section, let us consider a few explicit examples of strain configurations and Gr\"{u}neisen tensors. We start from the simplest case:
$\beta_{ij}^{(SS)}=\beta_{ij}^{(PP)}$ and $\beta_{x1}^{(x)}=\beta_{x2}^{(x)}=\beta_{z1}^{(x)}=\beta_{x1}^{(y)}=\beta_{x2}^{(y)}=\beta_{z1}^{(y)}=0$.
Under, such constraints, there is only the $z$ component of a strain-induced gauge potential
\begin{equation}
\label{results-revised-Azz-1-simple}
A_z^{(z)} = -\frac{2}{n_{\text{\tiny W}}ea} \sqrt{\frac{\beta}{M_0}} \left[\beta_{xx}^{(SS)}\left(u_{xx}+u_{yy}\right) +\beta_{xy}^{(SS)}\left(u_{yx}-u_{xy}\right) +\beta_{zz}^{(SS)}u_{zz}\right].
\end{equation}
Further, let us assume that the Gr\"{u}neisen tensors components satisfy the relations $\beta^{(SS)}_{ij}=\beta^{(PP)}_{ij}$,
$\beta_{x2}^{(y)}=\beta_{x1}^{(y)}=-\beta_{x1}^{(x)}$, and $\beta_{z3}^{(y)}=-\beta_{z3}^{(x)}$,
as well as consider the strains with $u_{zx}=u_{zy}=0$, $u_{xx}=u_{yy}$, and $u_{xy}=-u_{yx}$.
As one can easily check, in this case, $V_0 = V_x=V_y=0$ and
\begin{eqnarray}
\label{results-revised-Azz-1-simple-2}
A_z^{(z)} &=& -\frac{2}{n_{\text{\tiny W}} ea} \sqrt{\frac{\beta}{M_0}} \left[2\beta_{xx}^{(SS)}u_{xx} +2\beta_{xy}^{(SS)} u_{yx}  +\beta_{zz}^{(SS)}u_{zz}\right],\\
\label{results-revised-Axx-1-simple-2}
A_x^{(x)} &=& A_x^{(y)} =-\frac{2}{n_{\text{\tiny W}}ea} \sqrt{\frac{\beta}{M_0}} \beta_{z3}^{(x)} u_{xz},\\
\label{results-revised-Ayx-1-simple-2}
A_y^{(x)} &=& A_y^{(y)} =-\frac{2}{n_{\text{\tiny W}}ea} \sqrt{\frac{\beta}{M_0}} \beta_{z3}^{(x)} u_{yz}.
\end{eqnarray}
Therefore, there is an approximate analogy with the strain-induced gauge potential in Weyl semimetals. However, it is still incomplete
because the diagonal term with $\delta k_{\perp}^2$ [see the second term in Eq.~(\ref{results-revised-H-strain-1})] does not contain strain-induced fields at all.

\section{Optical lattice model}
\label{sec:Model-optical}

\subsection{General formulation}
\label{sec:Model-optical-general}

Since the effective solid-state Hamiltonian (\ref{results-revised-H-strain-1}) is rather
complicated, it is reasonable to analyze the effects of strains by using
a much simpler realization of the
double-Weyl semimetal phase in a noninteracting degenerate fermionic gas in an optical lattice.
In particular, we consider the following model of a 3D cubic optical lattice that contains double-Weyl nodes in its energy
spectrum \cite{Mai-Zhu:2017}:
\begin{eqnarray}
\label{model-optical-H}
H_{\rm OL} &=& t_0\sum_{\mathbf{r}} \sum_{j=x,y,z} \left[c^{\dag}_{\mathbf{r}+\mathbf{a}_j}U_{j}c_{\mathbf{r}} + c^{\dag}_{\mathbf{r}}U_{j}c_{\mathbf{r}+\mathbf{a}_j}\right]
- \frac{t_0}{4}\sum_{\mathbf{r}} \left[c^{\dag}_{\mathbf{r}+\mathbf{a}^{\prime}_1}\sigma_{y} c_{\mathbf{r}} +c^{\dag}_{\mathbf{r}}\sigma_{y}c_{\mathbf{r}+\mathbf{a}^{\prime}_1}\right]\nonumber\\
&+& \frac{t_0}{4}\sum_{\mathbf{r}} \left[c^{\dag}_{\mathbf{r}+\mathbf{a}^{\prime}_2}\sigma_{y}c_{\mathbf{r}} +c^{\dag}_{\mathbf{r}}\sigma_{y}c_{\mathbf{r}+\mathbf{a}^{\prime}_2}\right] +m_z \sigma_z \sum_{\mathbf{r}} c^{\dag}_{\mathbf{r}}c_{\mathbf{r}},
\end{eqnarray}
where $t_0$ is the hopping strength, $\mathbf{a}_j=a\hat{\mathbf{j}}$, $j=x,y,z$, $\mathbf{a}^{\prime}_1 = a\left(\hat{\mathbf{x}}+\hat{\mathbf{y}}\right)$, $\mathbf{a}^{\prime}_2 = a\left(\hat{\mathbf{x}}-\hat{\mathbf{y}}\right)$,
$m_z$ is the strength of an effective Zeeman potential,
\begin{eqnarray}
\label{model-optical-U-1}
U_x &=& \frac{\sigma_x-\sigma_z}{2},\\
\label{model-optical-U-2}
U_y &=& -\frac{\sigma_x+\sigma_z}{2},\\
\label{model-optical-U-3}
U_z &=& -\frac{\sigma_z}{2},
\end{eqnarray}
and $\sigma$-matrices act on the spin states of cold atoms.
In particular, while $\sigma_z$ corresponds to the spin-conserving hoppings, $\sigma_x$ and $\sigma_y$ represent the spin-flipping ones.
Note that the spin-flipping transitions can be realized by using the laser-assisted tunnelling technique with a specific Raman
coupling between two spin states \cite{Dalibard-Ohberg:2011,Galitski-Spielman:2013,Goldman-Spielman:2014,Zhai:2015}.

As expected in double-Weyl systems \cite{Fang-Bernevig:2012}, the lattice model (\ref{model-optical-H}) possesses the $C_4$ symmetry with
respect to the $z$ axis. Since we are interested in the effects of deformations on the low-energy electronic properties of the double-Weyl phases, we omit specific details of optical lattice realizations (for the corresponding details, see, e.g., Refs.~\cite{Dalibard-Ohberg:2011,Galitski-Spielman:2013,Goldman-Spielman:2014,Zhai:2015,Mai-Zhu:2017}).

By performing the Fourier transform
\begin{equation}
\label{model-optical-H-k-0}
H_{\rm OL}= \sum_{\mathbf{k}} c^{\dag}_{\mathbf{k}}\mathcal{H}_{\rm OL}(\mathbf{k})c_{\mathbf{k}},
\end{equation}
we obtain the following Hamiltonian of the optical lattice model in the momentum space:
\begin{equation}
\label{model-optical-H-k}
\mathcal{H}_{\rm OL}(\mathbf{k}) =  \sigma_z\left[m_z - t_0 \sum_{i=x,y,z} \cos{(ak_i)}\right] + \sigma_x t_0 \left[\cos{(ak_x)} -\cos{(ak_y)}\right] + \sigma_y t_0 \sin{(ak_x)} \sin{(ak_y)},
\end{equation}
As is easy to check, the double-Weyl phase with two double-Weyl nodes in the Brillouin zone is realized when $t_0<|m_z|<3t_0$.
For $|m_z|<t_0$ and $|m_z|>3t_0$, Hamiltonian (\ref{model-optical-H-k}) describes the topological and normal insulator phases, respectively \cite{Mai-Zhu:2017}. Henceforth, we will consider only the case of the double-Weyl phase.

The energy spectrum of Hamiltonian (\ref{model-optical-H-k}) reads
\begin{eqnarray}
\label{model-optical-energy}
\epsilon_{\mathbf{k}} &=& \pm \frac{1}{2}\Big\{4m_z^2 +11t_0^2 -8m_zt_0 \left[\cos{(ak_x)}+\cos{(ak_y)}+\cos{(ak_z)}\right] +3t_0^2 \cos{(2ak_y)} \nonumber\\
&+&t_0^2 \cos{(2ak_x)} \left[3+\cos{(2ak_y)}\right] +8t_0^2\cos{(ak_z)} \left[\cos{(ak_x)}+\cos{(ak_y)}\right]
+2t_0^2 \cos{(2ak_z)}\Big\}^{1/2}.
\end{eqnarray}
It contains two double-Weyl nodes located at $k_z=\pm b_z$, where $b_z=n_{\text{\tiny W}}/(2a)\arccos{\left(m_z/t_0-2\right)}$.

We present the energy spectrum (\ref{model-optical-energy}) for $m_z=2t_0$ at various values of momenta in
several panels of Fig.~\ref{fig:model-optical-energy-3D}. According to Fig.~\ref{fig:model-optical-energy-3D}(a),
the model possesses two double-Weyl nodes separated
by $2b_z$. Moreover, by comparing Figs.~\ref{fig:model-optical-energy-3D}(a) and
\ref{fig:model-energy-3D}(a), we find that there are no additional gapless features in the optical lattice
energy spectrum (\ref{model-optical-energy}) [cf. also Figs.~\ref{fig:model-optical-energy-3D}(b)
and \ref{fig:model-energy-3D}(b)]. The energy spectrum in the vicinity of a double-Weyl node
is shown in Fig.~\ref{fig:model-optical-energy-3D}(c) at $k_z=b_z$, where, as expected, the dispersion relation is quadratic.

\begin{figure}[t]
\begin{center}
\includegraphics[width=1\textwidth]{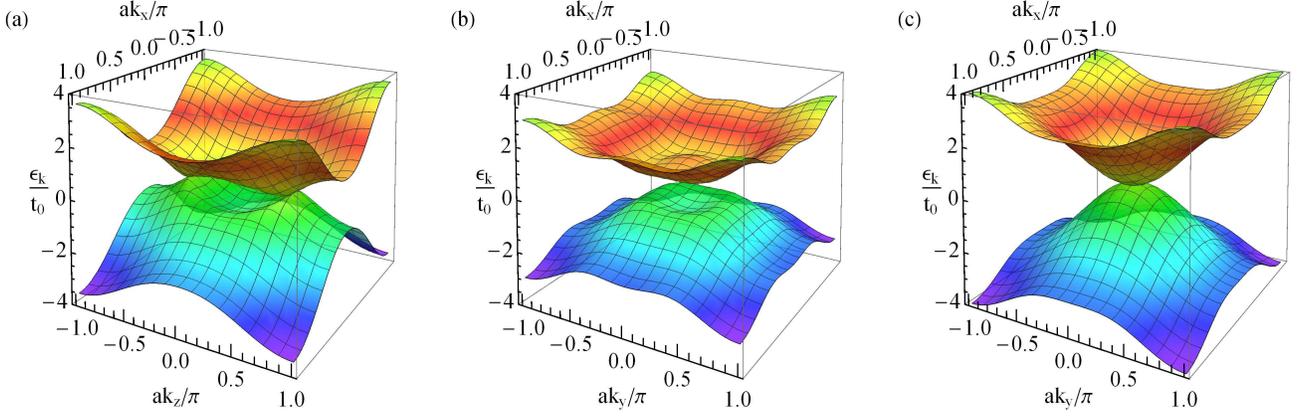}
\caption{The energy spectrum (\ref{model-optical-energy}) of the optical lattice model in the double-Weyl phase with $m_z = 2t_0$ at $k_y=0$ (panel a), $k_z=0$ (panel b), and $k_z=b_z$ (panel c).}
\label{fig:model-optical-energy-3D}
\end{center}
\end{figure}

Since we are interested in the low-energy properties of double-Weyl systems, it is convenient to expand Hamiltonian
(\ref{model-optical-H}) around the double-Weyl nodes in small deviations of momentum, i.e.,
$\mathbf{k}=\left\{\delta k_x, \delta k_y, \delta k_z +n_{\text{\tiny W}}/(2a) \arccos{\left(m_z/t_0-2\right)} \right\}$. We obtain
\begin{equation}
\label{model-optical-H-k-exp}
\mathcal{H}_{\rm OL}(\mathbf{k}) \approx \sigma_z \left[\frac{t_0a^2}{2} \delta k_{\perp}^2 +\frac{n_{\text{\tiny W}}a}{2} \sqrt{(3t_0-m_z)(m_z-t_0)} \delta k_z \right] -\frac{t_0a^2}{2} \left(\sigma_{+} \delta k_{+}^2 + \sigma_{-} \delta k_{-}^2\right) +O(\delta k_z^2,\delta k_z \delta k_{\perp}^2, \delta k_{\perp}^3).
\end{equation}
By comparing the linearized Hamiltonian in Eq.~(\ref{model-optical-H-k-exp}) with the model Hamiltonians in Ref.~\cite{Fang-Bernevig:2012}, it
is easy to see that for the small deviations from the Weyl nodes, the optical lattice model indeed possesses the minimal double-Weyl
structure.

\subsection{Strains in the optical lattice model}
\label{sec:strains-optical}

As in the case of the double-Weyl semimetals in Subsec.~\ref{sec:model-strains-2}, we include strains in the optical
lattice Hamiltonian (\ref{model-optical-H}) through the following modification of hopping parameters:
\begin{eqnarray}
\label{strains-optical-beta}
\sigma_z t_0 &\to& \sigma_z t_0\left[1 - \sum_{i=x,y,z} \beta_{ij}^{(z)} \frac{\left(\delta \mathbf{r}(\mathbf{a}_j) \cdot \mathbf{a}_i\right)}{|\mathbf{a}_i|}\right],\\
\label{strains-optical-beta-diag2}
\sigma_x t_0 &\to& \sigma_x t_0\left[1 - \sum_{i=x,y,z} \beta_{ij}^{(x)} \frac{\left(\delta \mathbf{r}(\mathbf{a}_j) \cdot \mathbf{a}_i\right)}{|\mathbf{a}_i|}\right],\\
\label{strains-optical-beta-diag3}
\sigma_y t_0 &\to& \sigma_y t_0\left[1 - \sum_{i=x,y,z} \beta_{in}^{(y)} \frac{\left(\delta \mathbf{r}(\mathbf{a}^{\prime}_n) \cdot\mathbf{a}_i\right)}{|\mathbf{a}_i|} \right],
\end{eqnarray}
where $j=x,y,z$, $n=1,2$, and the terms quadratic in $\delta r$ were omitted.

By using the $C_4$ symmetry of Hamiltonian (\ref{model-optical-H}) (see also Subsec.~\ref{sec:model-strains-2} for the corresponding discussion in the solid-state model), it is easy to simplify the Gr\"{u}neisen tensors.
We present the corresponding results in Eqs.~(\ref{strains-optical-beta-z-C4})--(\ref{strains-optical-beta-y-C4}) in Appendix \ref{sec:Gruneisen-app}.

Next, one can obtain the following linearized version of the strained Hamiltonian given by Eq.~(\ref{strains-optical-H-lattice-k}) in Appendix \ref{sec:H-app}:
\begin{eqnarray}
\label{strains-optical-H-k}
\mathcal{H}_{\rm OL}(\mathbf{k}) &\approx&
\sigma_z\left[\frac{t_0a^2}{2} \delta k_\perp^2  +\frac{n_{\text{\tiny W}}a}{2} \sqrt{(3t_0-m_z)(m_z-t_0)} \left(\delta k_z -eA_z\right)\right]
-\sigma_{+} \left(\frac{t_0a^2}{2}\delta k_{+}^2  +V_{+}\right) \nonumber\\
&-&\sigma_{-} \left(\frac{t_0a^2}{2}\delta k_{-}^2  +V_{-}\right) +O(\delta k_z^2, \delta k_z \delta k_{\perp}^2, \delta k_{\perp}^3, \delta k_z \hat{u},\delta k_{\perp}^2 \hat{u}),
\end{eqnarray}
where $V_{\pm} \equiv V_x \pm iV_y$ and
\begin{eqnarray}
\label{strains-optical-Vx}
V_x &=& at_0\left[\beta_{xx}^{(x)}\left(u_{xx}-u_{yy}\right) -\beta_{xy}^{(x)}\left(u_{xy}+u_{yx}\right)\right],\\
\label{strains-optical-Vy}
V_y &=& \frac{at_0}{2} \left[\beta_{x1}^{(y)}\left(u_{xx}+u_{xy}+u_{yx}-u_{yy}\right) -\beta_{x2}^{(y)}\left(u_{xx}-u_{xy}-u_{yx}-u_{yy}\right)\right], \\
\label{strains-optical-Az}
A_z &=& -\frac{2}{n_{\text{\tiny W}} e\sqrt{(3t_0-m_z)(m_z-t_0)}}
\left[t_0 \beta_{xx}^{(z)}\left(u_{xx}+u_{yy}\right)  -t_0\beta_{xy}^{(z)} \left(u_{xy}-u_{yx}\right) +(m_z-2t) \beta_{zz}^{(z)}u_{zz}\right].
\end{eqnarray}
Note that since both deformations and momenta deviations are small, we neglected
the high-order terms, i.e., $O(\delta k_z \hat{u},\delta k_{\perp}^2 \hat{u})$.
Physically, the latter provide only small modifications to the quasiparticles group velocity.

The energy spectrum of the linearized Hamiltonian (\ref{strains-optical-H-k}) reads
\begin{eqnarray}
\label{strains-optical-eps-k}
\epsilon_{\mathbf{k}} &=& \pm \Bigg\{
V_x^2+V_y^2+a^2t_0 \left[V_x\left(\delta k_x^2-\delta k_y^2\right) +2V_y \delta k_x \delta k_y\right] +\frac{a^4t_0^2}{4} \delta k_{\perp}^4 +\nonumber\\
&+& \left[\frac{a^2t_0}{2}\delta k_{\perp}^2 + \frac{n_{\text{\tiny W}} a}{2 }\sqrt{(3t_0-m_z)(m_z-t_0)} \left(\delta k_z -e A_z\right)\right]^2
\Bigg\}^{1/2}
\end{eqnarray}
and is shown at $\delta k_y=0$ in Fig.~\ref{fig:strains-optical-energy-3D-kx-kz} for $m_z =2t_0$, $A_z=0$, and several values of $V_x$ and $V_y$. In addition, we assumed that all parameters are uniform. Note that the strain-induced gap may resemble that in bilayer graphene \cite{Verberck-Trauzettel:2012}. (For a discussion of electronic properties in strained bilayer graphene, see, e.g., Ref.~\cite{McCann-Koshino:2013-rev}.)

\begin{figure}[t]
\begin{center}
\includegraphics[width=1\textwidth]{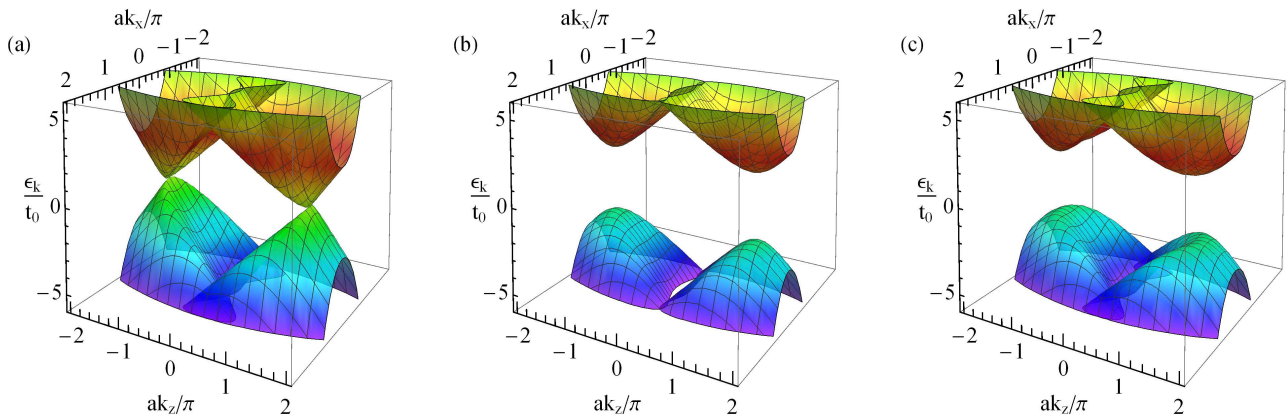}\hfill
\caption{The energy spectrum (\ref{strains-optical-eps-k}) at $A_z=V_x=V_y=0$ (panel a), $A_z=V_y=0$, $V_x=2t_0$ (panel b), and $A_z=V_x=0$, $V_y=2t_0$ (panel c) plotted for $m_z=2t_0$, $\delta k_y=0$, $k_x=\delta k_x$, and $k_z=n_{\text{\tiny W}} b_z/2+\delta k_z$.
}
\label{fig:strains-optical-energy-3D-kx-kz}
\end{center}
\end{figure}

It is worth noting that Hamiltonian (\ref{strains-optical-H-k}) supports the \emph{nematic} phase with a gapless energy spectrum.
This phase is characterized by an apolar ordering about the director $\mathbf{V}=\left\{V_x, V_y\right\}$, which means that the system has the symmetry $\mathbf{V} \leftrightarrow -\mathbf{V}$.
The possibility of a similar ground state was extensively discussed in the case
of bilayer graphene, see, e.g., Refs.~\cite{Vafek:2010,Lemonik:2010,Gorbar:2012ic}. Its
nematic order parameter can be achieved by applying strains~\cite{Mucha:2011,Son-Jhi:2011} and
rotational mismatch~\cite{Gail-Neto:2011} between the layers of bilayer graphene. We find
that the situation is slightly more complicated in 3D double-Weyl systems, where, unlike the 2D graphene, in order to get a nematic
phase, an additional condition should be satisfied.
In particular, the first term in
Eq.~(\ref{strains-optical-H-k}) should be set to zero, which leads to the following constraint for the
$z$ component of the momentum deviations:
\begin{equation}
\label{strains-optical-nematic-kz}
\delta k_z = eA_z - \frac{t_0a \delta k_\perp^2}{n_{\text{\tiny W}} \sqrt{(3t_0-m_z)(m_z-t_0)}}.
\end{equation}
For this value of $\delta k_z$, the energy spectrum (\ref{strains-optical-eps-k})
simplifies and reads
\begin{equation}
\label{strains-optical-nematic-eps-k}
\epsilon_{\mathbf{k}} = \pm \sqrt{|V|^2+a^2t_0 |V| |\delta k_{\perp}|^2 \cos{\left(2\varphi_k-\varphi\right)} +\frac{a^4 t_0^2}{4}|\delta k_{\perp}|^4},
\end{equation}
where we used $V_x=|V|\cos{\varphi}$, $V_y=|V|\sin{\varphi}$, $\delta k_x=|\delta k_{\perp}|\cos{\varphi_k}$, and $\delta k_y=|\delta k_{\perp}|\sin{\varphi_k}$.
As is clear from Eq.~(\ref{strains-optical-nematic-eps-k}), the expression under the square root is minimal in the two opposite directions given by $\varphi_k = \varphi/2 \pm \pi/2$ in the momentum space.
Indeed, in such a case, Eq.~(\ref{strains-optical-nematic-eps-k}) reads
\begin{equation}
\label{strains-optical-nematic-eps-k-1}
\epsilon_{\mathbf{k}} = \pm \left||V|-\frac{a^2t_0}{2} |\delta k_{\perp}|^2\right|,
\end{equation}
which vanishes at $|\delta k_{\perp}| =\sqrt{2|V|/(a^2t_0)}$.
The invariance of the dispersion relation (\ref{strains-optical-nematic-eps-k}) with respect to the transformations $\mathbf{V} \leftrightarrow -\mathbf{V}$ and $\varphi_k \to \varphi_k \pm \pi/2$ demonstrates that $\mathbf{V}$ is indeed a nematic order parameter in the system.

We present the energy spectrum (\ref{strains-optical-nematic-eps-k}) at $\delta k_z$ given by Eq.~(\ref{strains-optical-nematic-kz})
and $m_z =2t_0$ for various values of $V_x$ and $V_y$ in Fig.~\ref{fig:strains-optical-energy-3D-kx-ky-Nematic}.
As one can easily see, strains split a double-Weyl node with $n_{\text{\tiny W}}=2$ into two Weyl nodes each possessing a unit
topological charge.
As expected, the splitting depends on the relative contribution of the strain-induced terms $V_x$ and $V_y$.
It worth noting that the nematic phase is, in general, absent for the solid-state model (\ref{results-revised-H-strain-1}), where the effects of
strains are much more complicated. In passing, let us note that our findings agree with the results obtained in Refs.~\cite{Fang-Bernevig:2012,Huang,Mai-Zhu:2017}, where it was shown that $C_4$ symmetry breaking terms lead to the splitting of a double-Weyl node into a pair of Weyl nodes with the unit topological charge.
However, the possibility of the strain-induced nematic phase was not realised before.

\begin{figure}[t]
\begin{center}
\includegraphics[width=1\textwidth]{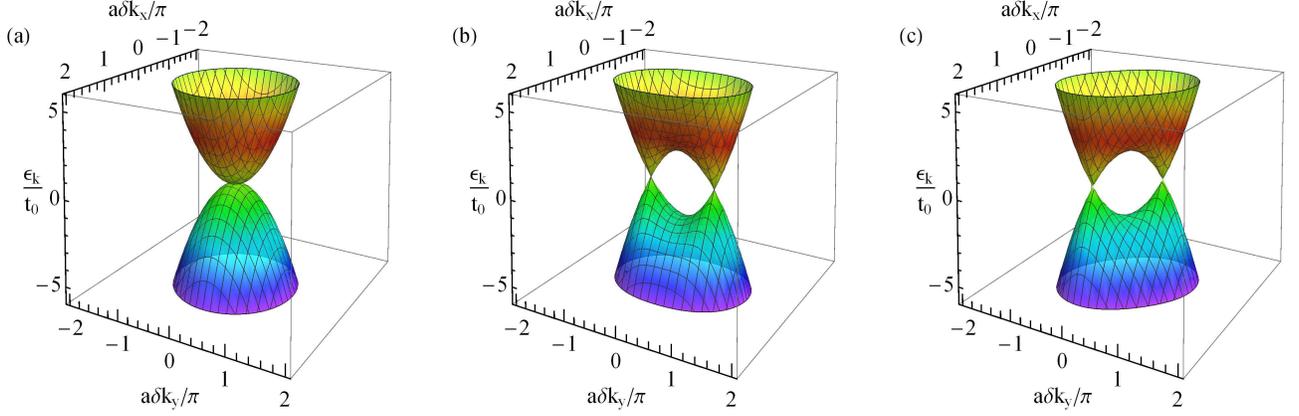}
\caption{The energy spectrum (\ref{strains-optical-nematic-eps-k}) at $V_x=V_y=0$ (panel a), $V_x=2t_0$, $V_y=0$ (panel b), and
$V_x=0$, $V_y=2t_0$ (panel c) plotted for $m_z = 2t_0$ and $\delta k_z$ given by Eq.~(\ref{strains-optical-nematic-kz}).
}
\label{fig:strains-optical-energy-3D-kx-ky-Nematic}
\end{center}
\end{figure}

\section{Wavepackets motion in strained optical lattice}
\label{sec:wavepackets}

In this section, to illustrate the effects of strains in double-Weyl phases, we analyse the quasiclassical motion of wavepackets in the deformed optical lattice model defined in Sec.~\ref{sec:Model-optical} (for a review of the
wavepackets dynamics in systems with the nontrivial Berry curvature, see Ref.~\cite{Xiao:2009rm}). Note also that the motion of the wavepackets in optical lattices with usual Weyl nodes was considered in Ref.~\cite{Roy-Grushin:2018}.
Since there are two double-Weyl nodes separated in momentum, it is reasonable to treat the evolution of the wavepackets from different Weyl nodes as independent.
Therefore, all variables that describe such a dynamics should have upper indices $(\pm)$ corresponding to the sign of the topological charge $n_{\text{\tiny W}}=\pm2$. Since we assume that the internode transitions are negligible, henceforth, such indices will be omitted.

\subsection{Equations of motion}
\label{sec:wavepackets-EOM}

A wavepacket centered at $\mathbf{r}(t)$ in the coordinate space with momentum $\mathbf{q}(t)$
is defined as a superposition of the Bloch states $\phi_{\mathbf{k}}=e^{i\mathbf{k}\mathbf{r}}\psi_{\mathbf{k}}$, i.e.,
\begin{equation}
\label{wavepackets-EOM-W-def}
W = \int \frac{d\mathbf{k}}{(2\pi)^3} a(t,\mathbf{k}) \phi_{\mathbf{k}}.
\end{equation}
Here $a(t,\mathbf{k})$ is a normalized distribution function that allows for the wavepackets localization at $\mathbf{r}(t)$ and $\mathbf{q}(t)$.
The eigenstates of the linearized Hamiltonian (\ref{strains-optical-H-k}) are given by
\begin{equation}
\psi_{\mathbf{k}} = N_{\mathbf{k}} \left\{-\frac{a^2t_0k_{\perp}^2 - a n_{\text{\tiny W}} \sqrt{(3t_0-m_z)(m_z-t_0)} (k_z-eA_z) +2\epsilon_{\mathbf{k}}}{a^2t_0k_{-}^2+2V_{-}}, 1\right\}^{\rm T},
\end{equation}
where $N_{\mathbf{k}}$ is the normalization constant.

The equations of motion for wavepackets in a weakly nonuniform clean medium with static strains are given by
\cite{Sundaram:1999zz,Panati-Teufel:2003,Shindou:2005vfm}
\begin{eqnarray}
\label{wavepackets-EOM-r-eq-def}
\dot{\mathbf{r}} &=& \mathbf{v}_{\mathbf{q}}(\mathbf{r},\mathbf{q}) +\hat{\Omega}_{\mathbf{q}\mathbf{r}}\dot{\mathbf{r}} +\hat{\Omega}_{\mathbf{q}\mathbf{q}}\dot{\mathbf{q}},\\
\label{wavepackets-EOM-q-eq-def}
\dot{\mathbf{q}} &=& -\mathbf{F}_{\mathbf{r}}(\mathbf{r},\mathbf{q}) -\hat{\Omega}_{\mathbf{r}\mathbf{r}}\dot{\mathbf{r}} -\hat{\Omega}_{\mathbf{r}\mathbf{q}}\dot{\mathbf{q}},
\end{eqnarray}
where
\begin{eqnarray}
\label{wavepackets-EOM-v-q-def}
\mathbf{v}_{\mathbf{q}}(\mathbf{r},\mathbf{q}) &=& \frac{1}{\hbar}\partial_{\mathbf{q}} \epsilon_{\mathbf{q}},\\
\label{wavepackets-EOM-v-r-def}
\mathbf{F}_{\mathbf{r}}(\mathbf{r},\mathbf{q}) &=& \frac{1}{\hbar} \partial_{\mathbf{r}} \epsilon_{\mathbf{q}}
\end{eqnarray}
are the components of the wavepackets group velocity and the effective force due to a lattice inhomogeneity, respectively.
Next, $\hat{\Omega}_{\mathbf{q}\mathbf{r}}\dot{\mathbf{r}}$ is a vector whose components are defined as
$\sum_{j=x,y,z}\left(\hat{\Omega}_{\mathbf{q}\mathbf{r}}\right)_{ij}\dot{r}_j$ and
the Berry curvature tensor $\hat{\Omega}_{\mathbf{q}\mathbf{r}}$ is
\begin{equation}
\label{wavepackets-EOM-Omega-qr-def}
\left(\hat{\Omega}_{\mathbf{q}\mathbf{r}}\right)_{ij} = -i\left[\left(\partial_{q_i}\psi_{\mathbf{q}}\right)^{\dag}\left(\partial_{r_j}\psi_{\mathbf{q}}\right) -\left(\partial_{q_j}\psi_{\mathbf{q}}\right)^{\dag}\left(\partial_{r_i}\psi_{\mathbf{q}}\right)\right].
\end{equation}
The expressions for the other products and tensors (i.e., $\hat{\Omega}_{\mathbf{q}\mathbf{q}}$, $\hat{\Omega}_{\mathbf{r}\mathbf{q}}$, $\hat{\Omega}_{\mathbf{r}\mathbf{r}}$) are obtained in a similar way.
Note that, in general, one needs to take into account a dissipative term on the right-hand side of Eq.~(\ref{wavepackets-EOM-q-eq-def}),
which describes the dissipation of wavepackets momenta. For the sake of simplicity, however, we ignore it.

\subsection{Wavepackets trajectories in the linearized model}
\label{sec:wavepackets-results-new}

In view of the complicated structure of Eqs.~(\ref{wavepackets-EOM-r-eq-def}) and (\ref{wavepackets-EOM-q-eq-def}) in the presence of strains,
we analyze the trajectories of the wavepackets numerically.
In addition, we consider small values of wavepackets momenta that justifies the use of the linearized model (\ref{strains-optical-H-k}).

For our numerical estimates, the following values of the lattice constant and the hopping strength \cite{Liu-Cheng:2013,Liu-Ng:2014,Wang-Duan:2014,Xu-Zhang:2014,Zhang-Wang:2016,Xu-Duan:2016} are employed:
\begin{equation}
\label{wavepackets-results-new-parameters-be}
a=764~\mbox{nm}, \quad t_0= 2.1\times10^{-12}~\mbox{eV}.
\end{equation}
Here the lattice constant $a$ corresponds to the blue-detuned laser wavelength
and we assume that
\begin{equation}
\label{wavepackets-results-new-beta}
 m_z=2t_0, \quad \beta_{ij}^{(z)}=\beta_{ij}^{(x)}=g/a, \quad \beta_{in}^{(y)}=g/a,
\end{equation}
where $g$ is a numerical coefficient, $i,j=x,y,z$, and $n=1,2$.
The initial values of the wavepackets positions and momenta are
\begin{eqnarray}
\label{wavepackets-results-new-initvals-r}
\mathbf{r}(t) &=& \mathbf{0},\\
\label{wavepackets-results-new-initvals-q}
\mathbf{q}(t) &=& 0.05 \frac{\pi}{a} \hat{\mathbf{y}}.
\end{eqnarray}
It is important to note that the lifetime of atoms in optical lattices can be as large as seconds
\cite{Liu-Cheng:2013} that makes possible to reliably track the motion of wavepackets for timescales up to tens of milliseconds.

For the sake of brevity, we consider the following two types of the double-Weyl optical lattices deformations: (i) the torsion of the optical lattice in the form of a wire about the $z$ axis and (ii) the bending of the thin optical lattice about the $y$ axis.
In the latter case, the unstrained lattice is located in the $x$-$y$ plane.

Let us start from the case of the torsion. The displacement vector is \cite{Landau:t7,Pikulin:2016,Arjona:2018ryu}
\begin{equation}
\label{wavepackets-results-new-torsion-u}
\mathbf{u} = \frac{\theta_{\rm twist}}{L} \left[\mathbf{r}\times \mathbf{z}\right],
\end{equation}
where $\theta_{\rm twist}$ denotes the total angle of the lattice twist and $L$ is the length along the torsion axis. The unsymmetrized strain tensor for the above displacement vector reads
\begin{equation}
\label{wavepackets-results-new-torsion-uij-z}
\hat{u} = \frac{\theta_{\rm twist}}{L} \left(
                   \begin{array}{ccc}
                     0 & -z & 0 \\
                     z & 0 & 0 \\
                     y & -x & 0 \\
                   \end{array}
                 \right).
\end{equation}
Then, as follows from Eqs.~(\ref{strains-optical-Vx})--(\ref{strains-optical-Az}), only $A_z$ is nonzero and equals
\begin{equation}
\label{strains-optical-models-torsion-Az}
A_z= -\frac{4 \theta_{\rm twist} t_0}{n_{\text{\tiny W}} eL\sqrt{(3t_0-m_z)(m_z-t_0)}}\beta_{xy}^{(z)} z.
\end{equation}
For our numerical estimates, we use the following parameters:
\begin{equation}
\label{wavepackets-results-new-parameters-torsion}
\theta_{\rm twist}=\pi, \quad L=10^4 a.
\end{equation}

Next, we consider the thin optical lattice bending about the $y$ axis.
While the undeformed lattice is located at the $x$-$y$ plane, the deformations lead to the following displacement vector \cite{Landau:t7,Arjona:2018ryu}:
\begin{equation}
\label{strains-optical-models-bending-u-y}
\mathbf{u} = \frac{u_0}{d} \left\{2xz,0,-x^2-D_{\rm Lame} z^2\right\}.
\end{equation}
Here $u_0$ defines the maximum stress, $d$ is the thickness of the optical lattice, and $D_{\rm Lame}$ is the relation between the Lam\'{e} coefficients \cite{Guinea-Novoselov:2010}. The unsymmetrized strain tensor for the above displacement vector is
\begin{equation}
\label{strains-optical-models-bending-uij-z}
\hat{u} = 2\frac{u_0}{d} \left(
                   \begin{array}{ccc}
                     z & 0 & -x \\
                     0 & 0 & 0 \\
                     x & 0 & -D_{\rm Lame} z \\
                   \end{array}
                 \right).
\end{equation}
Therefore, according to Eqs.~(\ref{strains-optical-Vx})--(\ref{strains-optical-Az}), the strain-induced terms $A_z$, $V_x$, and $V_y$ are
\begin{eqnarray}
\label{strains-optical-models-bending-Az}
A_z &=& - \frac{4 u_0}{n_{\text{\tiny W}} ed\sqrt{(3t_0-m_z)(m_z-t_0)}}\left[t_0\beta_{xx}^{(z)} -D_{\rm Lame} (m_z-2t)\beta_{zz}^{(z)}\right] z ,\\
\label{strains-optical-models-bending-Vx}
V_x &=& \frac{2u_0 at_0}{d} \beta_{xx}^{(x)} z,\\
\label{strains-optical-models-bending-Vy}
V_y &=& \frac{u_0 at_0}{d} \left(\beta_{x1}^{(y)} -\beta_{x2}^{(y)}\right) z.
\end{eqnarray}
It is reasonable to assume the following parameters of the bending:
\begin{equation}
\label{wavepackets-results-new-parameters-bending}
u_0=0.1, \quad d=50a, \quad D_{\rm Lame}=1.
\end{equation}

We present the projections of the wavepackets trajectories onto the $y$-$z$ plane in
Fig.~\ref{fig:wavepackets-results-new-torsion-trajectories-2D-g-01} for the two types of strains.
In the cases under consideration, the projections onto the other planes are trivial, i.e., the wavepackets move only in the $y$-$z$
plane. Since we use a low-energy approximation and neglect the internode scattering processes, the motion of the wavepackets from the
double-Weyl nodes with opposite topological charges can be considered as independent.

\begin{figure}[t]
\begin{center}
\includegraphics[width=1\textwidth]{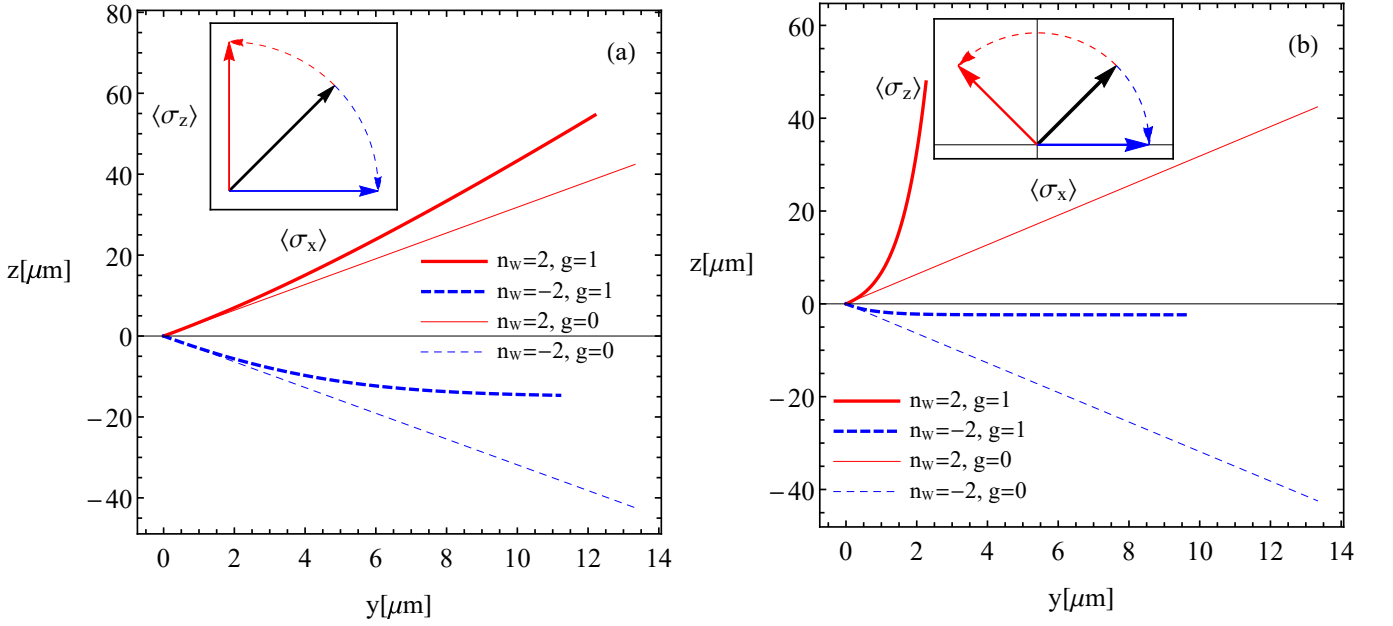}\hfill
\caption{The wavepackets trajectories projected onto the $y$-$z$ plane for the torsion about the $z$ axis (left panel) and the bending about the $y$ axis (right panel). The red solid and blue dashed lines correspond to $n_{\text{\tiny W}}=2$ and $n_{\text{\tiny W}}=-2$, respectively. The thick lines correspond to $g=1$ and the thin ones represent the undeformed case $g=0$.
The timescale is $t_{\rm max}=25~\mbox{ms}$.
Insets show the schematic illustrations of the spin projections evolution. While the black solid arrow represents the static spin orientation in the undeformed case, the red and blue solid arrows correspond to the asymptotic (at large $t$) spin orientations for the wavepackets with $n_{\text{\tiny W}}=2$ and $n_{\text{\tiny W}}=-2$, respectively, in the presence of deformations. Dashed arrows show the time evolution of spins.
}
\label{fig:wavepackets-results-new-torsion-trajectories-2D-g-01}
\end{center}
\end{figure}

As we can see from Fig.~\ref{fig:wavepackets-results-new-torsion-trajectories-2D-g-01}(a), the torsion about the $z$ axis breaks the mirror symmetry of the trajectories with respect to the $x$-$y$ plane and tends to either suppress ($n_{\text{\tiny W}}=-2$) or enhance ($n_{\text{\tiny W}}=2$) the $z$ component of the wavepackets velocity. Therefore, in essence, the torsion rotates the trajectories counterclockwise.
More interestingly, the wavepackets corresponding to opposite topological charges are spatially split even without deformations.
This phenomenon can be traced back to the dependence of the dispersion relation (\ref{strains-optical-eps-k}) and, consequently, the group velocity (\ref{wavepackets-EOM-v-q-def}) on the topological charge $n_{\text{\tiny W}}$.
Such a dependence appears due to the first term in Eq.~(\ref{strains-optical-H-k}).
Further, the amplitude of the spatial separation can be significant.
For example, at $q_y(t=0)=0.05\pi/a$, the splitting along the $z$ axis is almost an order of magnitude larger than the wavepackets path in the $y$ direction.
Such a difference is explained by the dispersion law (\ref{strains-optical-eps-k}), which
is quadratic in $\delta k_x$ and $\delta k_y$ but linear in $\delta k_z$ [see also Figs.~\ref{fig:model-optical-energy-3D}(a) and \ref{fig:strains-optical-energy-3D-kx-kz}(a)].
In such a case, the motion of the wavepackets is relativisticlike in the $z$ direction (with a finite at $\mathbf{q}\to\mathbf{0}$ group
velocity) and has a classical character for the $x$ and $y$ ones (with a linearly vanishing at $\mathbf{q}\to\mathbf{0}$ group
velocity). Therefore, when $q_x(t)$ or $q_y(t)$ is small, the wavepackets also move slowly in the corresponding direction. In is worth noting that the inhibition of the motion of the wavepackets from the different double-Weyl nodes is similar to the node-polarization effect predicted for a different setup with a usual Weyl semimetal in Ref.~\cite{Garrido-Munoz:2018}. However, while the latter effect is achieved by the combined effect of local torsion-induced pseudomagnetic and externally imposed magnetic fields, only the strains are present in our case.

Further, the results presented in Fig.~\ref{fig:wavepackets-results-new-torsion-trajectories-2D-g-01}(b) suggest that the bending also notably affects the motion of the wavepackets in a way qualitatively similar to the torsion. In particular, the $z$ component of the wavepackets velocity is clearly enhanced for $n_{\text{\tiny W}}=2$ and the propagation of the wavepackets with $n_{\text{\tiny W}}=-2$ is suppressed.
In order to check that the spatial splitting of the wavepackets from different double-Weyl nodes is not an artifact of the linearized model (\ref{strains-optical-H-k}), we also studied the wavepackets dynamics in the deformed lattice model given by Eq.~(\ref{strains-optical-H-lattice-k}) in Appendix \ref{sec:H-app}.
It is found that for the wavepackets with the initial momenta in the vicinity of the double-Weyl nodes, i.e., $q_z(t=0)= n_{\text{\tiny W}}/(2a) \arccos{\left(m_z/t_0-2\right)}$, the results are similar to those presented in
Fig.~\ref{fig:wavepackets-results-new-torsion-trajectories-2D-g-01}. This confirms that the strain-induced corrections $O(\delta q_z \hat{u},\delta q_{x}^2 \hat{u}, \delta q_{y}^2 \hat{u})$ neglected in Hamiltonian (\ref{strains-optical-H-k}) are indeed irrelevant for weak strains.

It is interesting to note that the mean value of the spin projection $\langle\bm{\sigma}\rangle$ also has a nontrivial dynamics.
For the initial conditions (\ref{wavepackets-results-new-initvals-r}) and (\ref{wavepackets-results-new-initvals-q}),
wavepackets from both double-Weyl nodes in the undeformed lattice have $\langle\sigma_x\rangle=\langle\sigma_z\rangle=1/\sqrt{2}$.
(Note that $\langle\sigma_y\rangle$ is always zero in the present setup.)
However, when the deformations are applied, the spin projections start to rotate in the $x$-$z$ plane. We present the schematic illustrations for the torsion and the bending in the corresponding insets in Fig.~\ref{fig:wavepackets-results-new-torsion-trajectories-2D-g-01}. As one can see from the inset in the left panel, the torsion allows for the complete spin polarization, albeit along the different axes ($z$ for the $n_{\text{\tiny W}}=2$ and $x$ for $n_{\text{\tiny W}}=-2$).
On the other hand, when the optical lattice is bent, one of the wavepackets remains unpolarized ($n_{\text{\tiny W}}=2$) and the other achieves polarization along the $x$ axis ($n_{\text{\tiny W}}=-2$). In passing, we note that the study of the spin polarization in double-Weyl optical lattices was also proposed in Ref.~\cite{Mai-Zhu:2017}. However, it is not related to the motion of wavepackets.

\section{Summary and discussions}
\label{sec:Summary}

In this study, we investigated the effects of strains on the low-energy dynamics in two models of a double-Weyl phase: a realistic solid-state model of a Weyl semimetal and a non-interacting fermionic gas in a 3D cubic optical lattice.
Deformations were taken into account via the change of the hopping parameters. The corresponding Gr\"{u}neisen tensors (i.e., the tensors that couple deformations to the hopping parameters) are constrained by the $C_4$ symmetry, which protects the double-Weyl nodes in both models.

It is found that, in both cases, strains do not couple to the low-energy sector only as an axial or, equivalently, a pseudoelectromagnetic gauge potential. The key to understanding such a difference from a usual Weyl semimetal is the structure of the corresponding low-energy Hamiltonians.
In particular, it is linear in momentum for the Weyl semimetals with the unit topological charge, i.e., $\propto\bm{\sigma}\cdot\mathbf{k}$.
Therefore, in general, perturbations, which do not depend on $\mathbf{k}$, only shift the positions of the Weyl nodes and, consequently, can be interpreted as a gauge potential.
This is clearly not the case in double-Weyl semimetals with the quadratic energy spectrum, where the similarity between strain-induced terms and a gauge potential can be established only in some special cases.
We expect that the same conclusion should be also valid for triple-Weyl semimetals whose band-crossing points have the topological charges $n_{\text{\tiny W}}=\pm 3$.

Our analysis of the optical lattice model with a simple structure provides a clear interpretation of deformation effects.
While there is a component of a pseudoelectromagnetic gauge potential in one directions, the
coupling in the other two, however, is of a different form.
What is more interesting, strains could lead to the formation of the \emph{nematic} phase when the $z$ component of momentum is fixed.
This phase is characterized by an apolar ordering about the strain-induced director.
Furthermore, in agreement with earlier symmetry-based findings~\cite{Fang-Bernevig:2012,Huang,Mai-Zhu:2017}, the double-Weyl nodes are split into pairs of nodes with the unit topological charges.

In order to illustrate the effects of strains on the electronic properties of the double-Weyl systems, we studied also the motion of wavepackets.
It is found that, even without strains, there is a clear separation of the wavepackets from the double-Weyl nodes with opposite
topological charges $n_{\text{\tiny W}}=\pm 2$.
Further, the torsion of the optical lattice wire with respect to the rotational symmetry axis affects the motion in the following way: it enhances the velocity of one of the wavepackets and suppresses the propagation of the other.
Additionally, there is a nontrivial dynamics of the spin projections in the optical lattice model.
In particular, the torsion leads to the complete spin polarizations of the wavepackets along different directions.

When a thin optical lattice is bent, it is found that the effects of strains are qualitatively similar to those in the case of the
torsion. In particular, depending on the topological charge, the wavepackets velocity can be either enhanced or suppressed.
Further, unlike the case of the torsion, only one of the wavepackets becomes completely spin-polarized.
Thus, while the separation of the wavepackets with opposite topological charges occurs even without strains, the latter qualitatively affect the trajectories and could lead to a spin polarization.

In passing, let us discuss the key limitations of this study. First, the explicit orbital structure of the states
in the solid-state model was not taken into account. While their composition can be complicated, it still
might provide some additional constraints on the Gr\"{u}neisen tensors. As for the optical lattice model,
the creation of sufficiently large lattices with strains similar to those in usual solids might be a difficult
task. Further, the realization of the nematic phase requires specific strain patterns that can be
nontrivial to control. In addition, the effects of interactions and disorder on the nematic order
parameter should be also taken into account. The corresponding investigations, however,
are outside the scope of this study.

\begin{acknowledgments}
The work of E.V.G. was partially supported by the Program of Fundamental Research of the
Physics and Astronomy Division of the National Academy of Sciences of Ukraine.
The work of V.A.M. and P.O.S. was supported by the Natural Sciences and Engineering Research Council of Canada.
The work of I.A.S. was supported by the U.S. National Science Foundation under Grants PHY-1404232
and PHY-1713950.
\end{acknowledgments}

\begin{center}
\textbf{Conflict of interest}
\end{center}

The authors have declared no conflict of interest.

\appendix

\section{Gr\"{u}neisen tensors and lattice Hamiltonians}
\label{sec:Gruneisen-H-app}

In this appendix, we present some technical details, such as the nonzero components of the Gr\"{u}neisen tensors and the Fourier transforms of the lattice Hamiltonians for both the solid-state and optical lattice models defined in Secs.~\ref{sec:model-solid} and \ref{sec:Model-optical}, respectively.

\subsection{Gr\"{u}neisen tensors components}
\label{sec:Gruneisen-app}

Let us start from the Gr\"{u}neisen tensors components.
In the case of the solid-state model (\ref{model-strains-2-H-r}) with the replacements (\ref{model-strains-2-tssy})--(\ref{model-strains-2-tspy}), the $C_4$ symmetry about the $z$ axis allows for the following nonzero components of the Gr\"{u}neisen tensors (see also Subsec.~\ref{sec:model-strains-2} in the main text):
\begin{eqnarray}
\label{model-strains-2-symmetry-beta-nonzero-be}
&&\beta_{zz}^{(SS)}, \quad \beta_{xx}^{(SS)} = \beta_{yy}^{(SS)}, \quad \beta_{xy}^{(SS)}=-\beta_{yx}^{(SS)},\\
&&\beta_{zz}^{(PP)}, \quad \beta_{xx}^{(PP)} = \beta_{yy}^{(PP)}, \quad \beta_{xy}^{(PP)}=-\beta_{yx}^{(PP)},\\
&&\beta_{x1}^{(x)} = -\beta_{x3}^{(x)}=\beta_{y2}^{(x)}=-\beta_{y4}^{(x)},\quad
\beta_{x2}^{(x)} = -\beta_{x4}^{(x)}=-\beta_{y1}^{(x)}=\beta_{y3}^{(x)}, \quad
\beta_{z1}^{(x)} = -\beta_{z2}^{(x)}=\beta_{z3}^{(x)}=-\beta_{z4}^{(x)},\\
&&\beta_{x1}^{(y)} = -\beta_{x4}^{(y)}=-\beta_{y2}^{(y)}=\beta_{y3}^{(y)}, \quad
\beta_{x2}^{(y)} = -\beta_{x3}^{(y)}=\beta_{y1}^{(y)}=-\beta_{y4}^{(y)}, \quad
-\beta_{z1}^{(y)} = \beta_{z2}^{(y)}=\beta_{z3}^{(y)}=-\beta_{z4}^{(y)}.
\label{model-strains-2-symmetry-beta-nonzero-ee}
\end{eqnarray}

Further, we consider the case of the double-Weyl optical lattices.
By using the $C_4$ symmetry of Hamiltonian (\ref{model-optical-H}) with replacements (\ref{strains-optical-beta})--(\ref{strains-optical-beta-diag3}), we find that the nonzero components of the Gr\"{u}neisen tensors are
\begin{eqnarray}
\label{strains-optical-beta-z-C4}
&&\beta_{zz}^{(z)}, \quad \beta_{xx}^{(z)} = \beta_{yy}^{(z)}, \quad \beta_{xy}^{(z)} = -\beta_{yx}^{(z)},\\
\label{strains-optical-beta-x-C4}
&&\beta_{zz}^{(x)}, \quad \beta_{xz}^{(x)}, \quad \beta_{yz}^{(x)}, \quad \beta_{xx}^{(x)} = \beta_{yy}^{(x)}, \quad \beta_{xy}^{(x)} = -\beta_{yx}^{(x)},\\
\label{strains-optical-beta-y-C4}
&&\beta_{y2}^{(y)} =\beta_{x1}^{(y)}, \quad \beta_{y1}^{(y)}=-\beta_{x2}^{(y)}.
\end{eqnarray}

\subsection{Fourier transform of the lattice Hamiltonians}
\label{sec:H-app}

In this subsection, we present the Fourier transforms of the lattice Hamiltonians for both solid-state and optical lattice models.
By using the lattice Hamiltonian in solids (\ref{model-strains-2-H-r}) with the replacements (\ref{model-strains-2-tssy})--(\ref{model-strains-2-tspy}) and performing the Fourier transform, we obtain the following Hamiltonian of a strained double-Weyl semimetal in the momentum space:
\begin{eqnarray}
\label{model-strains-2-H-k-strain}
\mathcal{H}(\mathbf{k})&=& -\frac{2\beta}{a^2} \sum_{i,j=x,y,z}\frac{\beta_{ij}^{(SS)}-\beta_{ij}^{(PP)}}{2} \frac{\left(\mathbf{a}_j\cdot\bm{\nabla}\right) \left(\mathbf{a}_i\cdot\mathbf{u}\right)}{|\mathbf{a}_i|} \cos{(\mathbf{k} \cdot\mathbf{a}_j)}
+\sigma_z \left(M_0-  \frac{6\beta}{a^2}\right) \nonumber\\
&+& \sigma_z \frac{2\beta}{a^2} \sum_{j=x,y,z}\left[1-\sum_{i=x,y,z}\frac{\beta_{ij}^{(SS)}+\beta_{ij}^{(PP)}}{2} \frac{\left(\mathbf{a}_j\cdot\bm{\nabla}\right) \left(\mathbf{a}_i\cdot\mathbf{u}\right)}{|\mathbf{a}_i|} \right] \cos{(\mathbf{k} \cdot\mathbf{a}_j)} \nonumber\\
&+&\sigma_x \frac{D}{a^3} \sum_{j=1}^4 \left[(-1)^{j+1} - \sum_{i=x,y,z}\beta_{ij}^{(x)} \frac{\left(\mathbf{a}_j^{\prime}\cdot\bm{\nabla}\right) \left(\mathbf{a}_i\cdot\mathbf{u}\right)}{|\mathbf{a}_i|} \right] \sin{(\mathbf{k} \cdot\mathbf{a}_j^{\prime})}  \nonumber\\
&-&\sigma_y \frac{D}{2a^3} \sum_{j=1}^4 \left[\left(2\delta_{j1}-1\right) - \sum_{i=x,y,z}\beta_{ij}^{(y)} \frac{\left(\mathbf{a}_j^{\prime\prime}\cdot\bm{\nabla}\right) \left(\mathbf{a}_i\cdot\mathbf{u}\right)}{|\mathbf{a}_i|} \right]  \sin{(\mathbf{k} \cdot\mathbf{a}_j^{\prime \prime})},
\end{eqnarray}
where $a$ is the lattice constant, $\mathbf{a}_j=a\hat{\mathbf{j}}$, $\hat{\mathbf{j}}$ denotes the unit vector
in the direction $j=x,y,z$, the lattice vectors $\mathbf{a}_n^{\prime}$ and $\mathbf{a}_n^{\prime \prime}$ with $n=\overline{1,4}$ are defined by Eqs.~(\ref{model-lattice-real-a1-1})--(\ref{model-lattice-real-a1-4}), the modification of the hopping length $\delta \mathbf{r}(\mathbf{a}_j)$ is given by Eq.~(\ref{model-strains-2-dr}) in the main text, $\mathbf{u}$ is the displacement vector, as well as $\beta$, $D$, and $M_0$ are the parameters of model (\ref{model-H-eff}). In addition, we assumed that coordinate dependence due to the strain-induced terms $\propto \partial_iu_j$ is weak and can be treated as a small spatial variation of the parameters in the momentum-space Hamiltonian.

The Fourier transform of the optical lattice Hamiltonian given in Eq.~(\ref{model-optical-H}) with deformations taken into account via Eqs.~(\ref{strains-optical-beta})--(\ref{strains-optical-beta-diag3}) takes the following form:
\begin{eqnarray}
\label{strains-optical-H-lattice-k}
\mathcal{H}_{\rm OL}(\mathbf{k}) &=& \sigma_z \Big\{m_z-t_0\cos{(ak_z)}\left[1-a\beta_{zz}^{(z)}u_{zz}\right] - t_0\cos{(ak_x)} \left[1-a\beta_{xx}^{(z)}u_{xx} +a\beta_{xy}^{(z)}u_{xy}\right] \nonumber\\
&-&t_0\cos{(ak_y)} \left[1-a\beta_{xx}^{(z)}u_{yy}-a\beta_{xy}^{(z)}u_{yx}\right] \Big\}
+t_0\sigma_{x}\Big\{\cos{(ak_x)}\left[1-a\beta_{xx}^{(x)} u_{xx} +a\beta_{xy}^{(x)} u_{xy}\right] \nonumber\\ &-&\cos{(ak_y)}\left[1-a\beta_{xx}^{(x)}u_{yy}-a\beta_{xy}^{(x)}u_{yx}\right]\Big\} \nonumber\\
&+&t_0\sigma_y\Big\{\sin{(ak_x)}\sin{(ak_y)}\left[1 -\frac{a}{2} \beta_{x1}^{(y)}\left(u_{xx} -u_{xy} +u_{yx}+u_{yy}\right) -\frac{a}{2}\beta_{x2}^{(y)}\left(u_{xx} +u_{xy} -u_{yx}+u_{yy}\right) \right] \nonumber\\
&+&\cos{(ak_x)}\cos{(ak_y)}\frac{a}{2} \left[\beta_{x1}^{(y)}\left(u_{xx} +u_{xy} +u_{yx}-u_{yy}\right) -\beta_{x2}^{(y)}\left(u_{xx} -u_{xy} -u_{yx}-u_{yy}\right)\right]\Big\},
\end{eqnarray}
where $t_0$ is the hopping strength, $m_z$ is the strength of an effective Zeeman potential, and unsymmetrized strain tensor is $u_{ij}=\partial_{i}u_j$, where $i,j=x,y,z$.
As in the case of the solid-state Hamiltonian (\ref{model-strains-2-H-k-strain}), it is assumed that the coordinate dependence of the unsymmetrized strain tensor is weak and provides a small spatial modulation of the Hamiltonian parameters. In addition, we used the results in Eqs.~(\ref{strains-optical-beta-z-C4})--(\ref{strains-optical-beta-y-C4}).

\end{document}